\documentclass[12pt]{article}
\usepackage{epsfig}
\begin{document}
\begin{center}
\Large
{\bf Exchange Currents in
Photoproduction of Baryon Resonances}
\normalsize
\end{center}
\begin{center}
U. Meyer\footnote{Supported by the DFG Graduiertenkolleg MU 705/3}, A.J. Buchmann, and Amand Faessler\\[12pt]
{\it Institute for Theoretical Physics, University of T\"{u}bingen\\
Auf der Morgenstelle 14, D--72076 T\"{u}bingen, Germany}\\
\end{center}
\begin{center}
{\bf Abstract}
\end{center}
\begin{small}
We calculate photoexcitation amplitudes for all nucleon and 
$\Delta$ resonances up to $\sqrt{s}$=1.6 GeV.
 We use a chiral quark model including two-body exchange 
currents. The two-body currents give important contributions.
 For the $\Delta$(1232) and the D$_{13}(1520)$ we observe that the individual 
exchange current contributions considerably cancel
each other while for the P$_{11}$(1440), the P$_{33}$(1600), and 
the S$_{11}$(1535) we get a reinforcement of the two-body amplitudes.
 In comparison with present experimental data, we obtain both for 
the S$_{11}$(1535) and for the P$_{11}(1440)$ amplitudes an improvement
with respect to the impulse approximation.
\end{small}\\[24pt]
\newpage
The study of electro-and photocouplings of nucleon resonances gives insight
into the internal structure of these resonances and is therefore important
for our understanding of low-energy quark dynamics. The simultaneous
description of the baryon mass spectrum and of 
electromagnetic (e.m.) transition
amplitudes is certainly a more severe test for the quark-quark interaction
than a description of the mass spectrum alone.\par
 From the experimental side, with the construction of continuous wave electron
accelerators, new and more exact data for nucleon resonance masses,
widths, and photocouplings are expected to be obtained in the next years.
 The accuracy and statistics of these data will be comparable to those
of hadronic processes \cite{Burkert}.\par
From the theoretical side, the e.m.\ excitation of the nucleon 
has mostly been studied in the quark model [2-9]. We also use in the
present calculations a Chiral Quark Potential Model ($\chi$QPM).
 In the simple nonrelativistic quark model [3,5] the description 
of photoexcitation amplitudes has only partly 
been satisfactory. One reason might be the neglect of relativistic
corrections in the single-quark current 
operator. These have been investigated e.g.\ in Refs.[4,7,8] and
have in some cases led to a better agreement with the data. Another
reason might be that most calculations have been done in the so-called
impulse approximation where the whole photon four-momentum is transferred
to one single constituent of the baryon, the other two quarks being spectators in the photon absorption process. However, in order to formulate
a gauge invariant model, it has been
known for a long time that it is necessary to take into account two-body
exchange currents \cite{Siegert}.\par
Two-body exchange currents have been shown to be important for
quantities such as the neutron charge radius \cite{BHY} 
or the magnetic moments of the nucleon \cite{Alf3} and 
the entire
baryon octet \cite{Georg}. In addition, they 
play an important role for the photoexcitation of
the two lightest resonances, namely
the $\Delta$-isobar ($P_{33}(1232)$) \cite{Alf1} and the Roper 
resonance ($P_{11}(1440)$) \cite{Uli}.
 We found that for the Roper resonance the inclusion of  
two-body currents slightly improves the agreement 
with experimental data. A similar improvement does not occur for 
the M1 transition amplitude of the $\Delta$-isobar
where the various exchange current contributions nearly 
completely cancel each other [14,15].
 In this work we extend our study of the effect of two-body currents on
photoproduction amplitudes of nucleon resonances by considering all 
excitations up to $\sqrt{s}$=1.6 GeV in order to investigate further
whether
the differences between quark model calculations and
experimental values arise from the non-completeness of the e.m.\ current,
 i.e. from the neglect of two-body contributions.\par
The Hamiltonian of the $\chi$QPM is in
the case of equal constituent quark masses $m_q$ given by \cite{Georg}
\begin{equation}
H=\sum_{i=1}^3(m_q+\frac{{\bf p}_i^2}{2m_q})-\frac{{\bf P}^2}{6m_q}
+\sum_{i<j}^{3}(V^{conf}({\bf r}_i,{\bf r}_j)+
V^{OPEP}({\bf r}_i,{\bf r}_j)+
V^{OSEP}({\bf r}_i,{\bf r}_j)+
V^{OGEP}({\bf r}_i,{\bf r}_j))\hspace*{2mm}.
\end{equation}
Here, ${\bf p}_i$(${\bf r}_i$) describes the momentum (position)
of the $i$-th constituent and ${\bf P}$
the center of mass momentum, respectively. The interaction between the
constituent quarks consists of various terms,
which model the main symmetries and
the dynamical content of QCD in the 
low-energy region.
The spontaneous breakdown of chiral symmetry leads to 
the appearance of the pion as Goldstone boson and 
its chiral partner, the sigma meson. 
The coupling of these mesons to the
constituent quarks is described in lowest order by 
the one-pion- ($V^{OPEP}$) and the one-sigma meson ($V^{OSEP}$) exchange 
potentials. 
The short-range part of the quark-quark interaction is mainly described
by a one-gluon exchange potential ($V^{OGEP}$). 
Finally, we model the quark confinement by a 
phenomenological two-body harmonic oscillator potential ($V^{conf}$).
Explicit expressions
for the interaction terms may be found for example in Ref.\cite{Georg}.\par
For the orbital part of the 
wavefunctions we use unmixed harmonic oscillator eigenfunctions. Explicit 
expressions for the orbital wavefunctions are given in Ref.\cite{Giannini}.\par
In the $\chi$QPM we have six independent parameters. These are the 
constituent quark mass $m_q$, the quark-gluon coupling constant $\alpha_s$,
the oscillator parameter $b$ and the confinement strength $a_c$.
 In addition, one should think of constituent quarks as extended particles
with finite hadronic and electromagnetic size. Therefore,
 we introduce a cut-off parameter $\Lambda$ 
in the one-pion and one-sigma meson
exchange potentials. This cut-off results in extended quark-pion and
quark-sigma meson vertices and can be interpreted
 in terms of a finite hadronic size of the
quarks. The finite e.m.\ size of the constituents $r_{\gamma q}$ is 
parametrized by a monopole form factor which is multiplied with the 
one- and two-body current densities as described 
in Refs.\cite{Alf3} and \cite{Georg}. We assume that the pion-photon
and quark-photon vertices are parametrized by one and the same monopole
form factor as discussed in Ref.\cite{Alf1}. 
In this way, the continuity equation for the e.m.\ current is still 
fulfilled.\par
 The set of parameters used here is listed in Table 1. We chose a value of
$r_{\gamma q}$=0.6 fm which is compatible with a prediction of the
simplest vector meson dominance model \cite{Weise}.
 The way how the other parameters are determined is explained
in Ref.\cite{Alf3}.\par 
\hspace*{-0.8cm}\marginpar{Table 1}\\
The coupling of the pion and the sigma meson to the quarks and the 
sigma meson mass are fixed
according to the chiral symmetry arguments of Refs.\cite{Obu} and \cite{Fae}:
\begin{equation}
\frac{g_{\sigma q}^2}{4\pi}=\frac{g_{\pi q}^2}{4\pi}=\left(\frac{3}{5}
\frac{m_q}{m_N}\right)^2 \frac{g_{\pi N}^2}{4\pi},\hspace*{1cm}
m_{\sigma}^2=4m_q^2+\mu^2\hspace*{2mm}.
\end{equation}
Here $m_N=939$ MeV is the nucleon mass and $\mu=138$ MeV the pion mass.
 Furthermore, we chose $g_{\pi N}^2/4\pi=13.845$.\par
In order to calculate electromagnetic properties of baryons, we 
have to know the four-vector current density $j_\mu=(\rho(x),
-{\bf j}(x))$ of the interacting quarks in the baryon system. In the past,
 most calculations were done in the so-called impulse approximation,
 where a pure one-body current was used and two-body currents were neglected. 
 However, a baryon is a strongly correlated system
with strong interactions between the three constituents. Thus, it is
intuitively clear that this simple 
description of the photon absorption
process is incomplete. In fact, it has been shown that two-body
currents are necessary to guarantee electromagnetic current
conservation \cite{Alf3}. The complete current density ({\bf j}$_{tot}$) is 
therefore given by a sum of one-and two-body operators:
\begin{equation}\label{curr}
{\bf j}_{tot}({\bf q})=\sum_{i=1}^{3}{\bf j}_{imp}({\bf r}_i)+
\sum_{i<j=1}^{3}(
{\bf j}_{g\overline{q}q}({\bf r}_i,{\bf r}_j)+
{\bf j}_{\pi\overline{q}q}({\bf r}_i,{\bf r}_j)+
{\bf j}_{\gamma\pi\pi}({\bf r}_i,{\bf r}_j)+
{\bf j}_{conf}({\bf r}_i,{\bf r}_j)+
{\bf j}_{\sigma\overline{q}q}({\bf r}_i,{\bf r}_j))\hspace*{2mm},
\end{equation}
where ${\bf j}_{imp}({\bf r}_i)$ is the one-body current density operator and
${\bf j}_{g\overline{q}q}$ (${\bf j}_{\pi\overline{q}q}$) corresponds to 
the current density due to one-gluon-(one-pion) exchange.
In addition, we obtain currents from the scalar part of the interaction,
 namely from the confinement potential (${\bf j}_{conf}$) and the one-sigma
meson exchange (${\bf j}_{\sigma\overline{q}q}$). For the pion we 
get an additional part
 (${\bf j}_{\gamma\pi\pi}$) from the pion-in flight diagram (Fig.1(c))
describing the photon coupling to the exchanged
pion. The one -and two-body 
current densities may be obtained by a nonrelativistic expansion of the 
Feynman amplitudes of the diagrams displayed in Fig.1. Alternatively, the 
various contributions may be calculated by
minimal substitution in the corresponding interaction terms of the 
$\chi$QPM Hamiltonian. Explicit expressions for the one- and two-body operators
may be found in Ref.\cite{Alf3}. 
 Note, that the parameters for the present calculation of 
electromagnetic excitation amplitudes are the same as the ones used for the 
calculations of the e.m.\ properties of the nucleon and $\Delta$(1232) [13-15]
and that we have not introduced any further parameters.\par
\vspace*{2pt}\marginpar{Figure 1}
The electromagnetic excitation of the nucleon is determined by the
photon-baryon interaction Hamiltonian
\begin{equation}
H_{em}=\int d^4 x J_{\mu}(x)A^{\mu}(x)\hspace*{2mm}.
\end{equation}
The quantities which are usually extracted 
from photo pionproduction experiments are the so-called
helicity amplitudes. For the excitation with real photons we only
need the transverse helicity amplitudes defined by \cite{Copley}
\begin{equation}
A_{\lambda}^{N}=-e\sqrt{\frac{2\pi}{\omega}}
\langle B^*, \lambda\mid\mbox{\boldmath$\epsilon$}\cdot{\bf j}({\bf q})\mid N,\lambda-1\rangle\hspace*{2mm}.
\end{equation}
They describe the transition of the nucleon with total angular momentum
projection $J_z=\lambda-1$
to a resonance with total angular momentum projection $J_z=\lambda$ 
through the absorption of a photon with positive helicity.
 Here, $\omega$ is the energy transfer of the photon in the center of mass
frame and $\mbox{\boldmath$\epsilon$}$ is the photon polarization.\par
The transverse helicity amplitudes may be expressed in terms of electromagnetic multipole transition
amplitudes. For this one decomposes the three-vector current
density into electric and magnetic multipole operators 
\cite{ForestWalecka},
\begin{equation}
j_{m}({\bf q})= -\sqrt{2\pi}\sum_{J=1}^{\infty}i^J\sqrt{2J+1}\left[m
T^{[M]J}_{{}{}m}(q)+T^{[E]J}_{{}{}m}(q)\right]
\end{equation}
with $m=0,\pm 1$ being the three-vector current density components
in the spherical basis. Because of the selection rules due to parity
and angular momentum conservation, only a few multipole operators
contribute. The
$J=\frac{3}{2}$ resonances with positive parity, namely the 
P$_{33}$(1232) and P$_{33}$(1600), are excited by M1 and 
E2 multipoles while the P$_{11}$(1400) is excited by M1 radiation only. 
 Similarly, for the negative parity excitations the S$_{11}$(1535) 
is excited by
E1 radiation while the D$_{13}$(1520) resonance can be reached by
E1 and M2 transitions.\par
In Table 2, we give
the results for the photocouplings of the positive parity resonances of the 
nucleon. For the helicity amplitudes, we use the same phase convention 
as Koniuk and Isgur \cite{Koniuk}. The P$_{33}$(1232) and the P$_{11}(1440)$ 
excitations have already been discussed in previous 
works [14,15], but now the helicity amplitudes
are calculated for finite $|{\bf q}|=\omega$. Therefore,
 the e.m.\ size of the constituent quarks contributes. The
finite e.m.\ size of the constituent quarks has
already been shown to be important for the proton charge radius \cite{Alf3} 
and for the magnetic radii of the baryon octet [12,13].
 Furthermore, we show in Table 2 our results for the  
P$_{33}$(1600) resonance. For the $\Delta$-isobar and 
its orbital excitation P$_{33}$(1600), the results for
the proton and neutron helicity amplitudes are identically the same.
We see that for all resonances the contributions from the 
two-body currents are important. This especially holds for the confinement
current which in some cases gives a 60$\%$-contribution relative to the
one-body current.
 In comparison with our results in
Ref.\cite{Alf1} and \cite{Uli}, we obtain smaller excitation amplitudes 
due to the finite e.m.\ size of the constituents.
Thus, for the $\Delta$-isobar we are not 
able to improve our previous results [14,15]
which deviate by 30-40$\%$ from
experimental values. This difference is even slightly enlarged by
the two-body currents and the finite e.m.\ size of the quarks. Another 
difference 
compared with our previous calculations in Refs.\cite{Alf1} and
\cite{Uli} arises from the different treatment of the 
E2 transition amplitude. In Refs.\cite{Alf1} and \cite{Uli} we 
calculated the E2 amplitude with
the charge density in the long-wavelength limit \cite{Siegert}
which leads to a relation between transverse electric and 
Coulomb transition amplitudes. Here, however, the E2 amplitude is
calculated using the current density, which yields a vanishing
E2 amplitude with unmixed harmonic oscillator wavefunctions. \par
In contrast to the $\Delta$(1232), for the Roper resonance, as 
already discussed in \cite{Uli},
 the two-body contribution tends in the right direction. However, in 
our previous calculations \cite{Uli} the total amplitude came out too large. This
is now improved by taking into account
the finite e.m.\ size of the constituent quarks. For the Roper resonance,
 by virtue of the two-body contributions and the finite e.m.\ size, our new 
results are 
in better agreement with the experimental values
than the results of the pure impulse approximation are.\par
 As in the case of the Roper resonance, the exchange current
contributions to the P$_{33}$(1600) resonance reinforce each other. The 
P$_{33}$(1600) excitation is only a ($\star\star\star$)-resonance and 
the experimental data are less reliable. We obtain
a good description of the $A_{1/2}$ amplitudes, but for the $A_{3/2}$ 
amplitudes a major difference between the calculated and experimental 
values, which is even enlarged by the two-body contributions.\par
\vspace*{2pt}\marginpar{Table 2}
Looking at the negative parity resonances (Table 3), we again find that
the two-body currents and especially the confinement current give
important contributions to the helicity amplitudes.
 For the S$_{11}(1535)$ we obtain
that the individual two-body contributions add 
constructively. For the neutron excitation, we get after including 
the two-body currents a similar good agreement with the experimental value as 
in impulse approximation. For the proton target, we obtain the 
interesting
result that the helicity amplitude of the proton excitation to the 
S$_{11}(1535)$ resonance is strongly damped by the two-body amplitudes.
 Thus, after adding two-body operators to the current density 
the long-standing problem that nonrelativistic quark model calculations
considerably overestimate the proton excitation
to the S$_{11}(1535)$ resonance \cite{Koniuk}, disappears. Our result 
is now in good agreement with experimental
values from Ref.\cite{PDG96}.
\par
For the D$_{13}(1520)$ excitation amplitudes, we observe, in contrast
to the S$_{11}(1535)$, 
 at least for the $A_{3/2}$ amplitude cancellations of the 
individual two-body amplitudes. For the 
$A_{1/2}$ amplitude of the proton, the agreement with experiment is 
improved whereas the difference between the calculated $A_{3/2}$ amplitudes 
and the experimental values cannot be reduced.\par 
 We have listed in Tables 2 and 3 the contributions from the
different multipole operators separately. For the $\Delta$-isobar and
the P$_{33}(1600)$, the E2 amplitude is zero (Table 2).
 For the D$_{13}$-excitation (Table 3), neither
the E1 nor the M2 amplitude dominates the 
photoexcitation amplitudes. In order to further investigate the 
remaining differences
between experimental extractions and our calculations 
of the $A_{3/2}$ amplitudes for the D$_{13}$ excitation and of
corresponding amplitudes for the $\Delta$-isobar, a detailed 
study of the $\chi$QPM Hamiltonian with
perturbed harmonic oscillator wavefunctions should be made.\par
\vspace*{2pt}\marginpar{Table 3}
Finally, we would like to concentrate on the pion-in-flight 
contribution (Fig.1(c)) whose
importance has been discussed by several groups, recently.
 First, Robson \cite{Robson} considered the effect of pion exchange
currents in the photoproduction of several resonances. For the pion
pair contribution, we agree with his calculations for the positive parity
resonances if we consider
 a zero e.m.\ size of the constituents and a pointlike quark-pion vertex.
 However, Robson claimed that the pion-in-flight contributions 
are not significant for all excitations he studied and 
neglected them. This opinion is later joined by Perazzi 
et.al. in a new calculation \cite{Boffi}.
 In contrast, we get for the positive 
parity resonances important contributions from the pion-in-flight
term. For the positive parity
excitations, the pion-in-flight amplitude is even larger than the one from the
pion-pair current. However, the two contributions from the pion
are of different sign and therefore the total effect of the pion 
as the sum of both is comparatively small, as has already been 
pointed out in Ref.\cite{Alf1} for the 
case of the $\Delta$. Therefore,
 previous calculations [9,22] that neglected 
the pion-in-flight term 
for positive parity excitations considerably overestimated the pion 
exchange current contributions. For the negative parity resonances,
 the amplitude from the pion-in-flight current is not as large as the 
one from the pion pair current, but again it gives, at least with 
our usual set of parameters of Table 1,
 non-negligible contributions.
 Looking at the contributions of different multipole 
operators to the total pion-in-flight amplitude, we 
realize that for the P$_{33}$(1232) and the P$_{33}$(1600) the
total amplitude is given by the M1 multipole operator whereas for 
the D$_{13}$(1520) only the E1 operator contributes. The E2 and M2 amplitudes
vanish due to the spin-isospin structure of the wavefunctions. Therefore, one 
has to properly take into account the pion-in-flight contributions.\par 
 Summarizing, we calculated the photoexcitation
amplitudes of all nucleon resonances
up to $\sqrt{s}=1.6$ GeV.
 We observed that exchange current contributions, and especially the
confinement current, give significant 
contributions for the helicity amplitudes
of all nucleon resonances. For the P$_{11}(1440)$ 
and the S$_{11}(1535)$, we get an 
improvement with respect to the impulse 
approximation. This, however, does %
not hold as a general tendency.
 For example, for the P$_{33}(1232)$ or the D$_{13}(1520)$, we are 
not able to get a better agreement with the data. \par 
In a future work, the baryon orbital wavefunctions should
be expanded in a larger Hilbert space. In addition, due to 
its
importance for the e.m.\ production of nucleon resonances, the 
phenomenological confinement potential should
be examined in more detail. In particular, different radial forms
(linear, color-screened, $\ldots$) should be studied concerning
their effect both on the baryon spectrum and e.m.\ transition amplitudes.\\[12pt]
{\bf Acknowledgment}: We would like to thank B. Krusche and W. Pfeil 
for fruitful discussions. 
\newpage
\begin{center}
\begin{tabular}{|r|r|r|r|r|r|}
\hline
$m_q$ [MeV]&$b$ [fm]&$\alpha_s$&$a_c$ [MeV$\cdot$fm$^{-2}]$&
$\Lambda$ [fm$^{-1}$]&$r_{\gamma q}$ [fm]\\
\hline
313&0.613&1.093&20.92&4.2&0.6\\
\hline
\end{tabular}\\[4pt]
{\small Table 1: Quark model parameters}\\[24pt]
\begin{tabular}{|c|r|c|c|c|c|c|c||c|c|}
\hline
{}&{}&$A_{i}$&$A_{g}$&$A_{\pi q\bar q}$&$A_{\gamma\pi\pi}$&$A_{c}$&
$A_{\sigma}$&$A_{tot}$&exp. \cite{PDG96}\\
\hline
$P_{33}$(1232)&$A_{1/2}(M1)$&--94&--10&+15&--18&+36&--11&--82&{}\\
{}&$A_{1/2}(E2)$&0&0&0&0&0&0&0&{}\\
{}&$A_{1/2}$(tot)&--94&--10&+15&--18&+36&--11&--82&--140$\pm$5\\
\cline{2-10}
{}&$A_{3/2}(M1)$&--163&--17&+26&--31&+62&--19&--142&{}\\
{}&$A_{3/2}(E2)$&0&0&0&0&0&0&0&{}\\
{}&$A_{3/2}$(tot)&--163&--17&+26&--31&+62&--19&--142&--258$\pm$6\\
\hline
$P_{11}$(1440)&$A_{1/2}^p$(M1)&--30&--16&+5&--13&--21&--15&--90&--65$\pm$4\\
{}&$A_{1/2}^n$(M1)&+20&+5&--8&+13&+14&+10&+54&+40$\pm$10\\
\hline
$P_{33}$(1600)&$A_{1/2}(M1)$&--19&--4&+5&--8&--4&--7&--37&{}\\
{}&$A_{1/2}(E2)$&0&0&0&0&0&0&0&{}\\
{}&$A_{1/2}$(tot)&--19&--4&+5&--8&--4&--7&--37&--23$\pm$20\\
\cline{2-10}
{}&$A_{3/2}(M1)$&--33&--7&+8&--14&--6&--12&--64&{}\\
{}&$A_{3/2}(E2)$&0&0&0&0&0&0&0&{}\\
{}&$A_{3/2}$(tot)&--33&--7&+8&--14&--6&--12&--64&--9$\pm$21\\
\hline
\end{tabular}\\[2pt]
\begin{minipage}[b]{14.5cm}
{\small Table 2: Helicity amplitudes for the $\gamma N\rightarrow P_{33}(1232),
P_{11}(1440), P_{33}(1600)$ transitions in units
of GeV$^{-1/2}\times 10^{-3}$ with the parameters of Table 1
evaluated at $|{\bf q}_{c.m.}|=\omega_{c.m.}$. $A_{i}$ = impulse, $A_{g}$ = gluon, 
$A_{\pi q\bar q}$ = pion pair, $A_{\gamma\pi\pi}$ = pion-in-flight, $A_{c}$ = 
confinement, and $A_{\sigma}$ = sigma meson contributions. $A_{tot}$ 
is the sum of all contributions. The proton and neutron transition 
amplitudes to the $P_{33}(1232)$ and $P_{33}(1600)$ resonances are 
the same. The M1 and E2 contributions are given separately.}
\end{minipage}\\[24pt]
\end{center}
\newpage
\begin{center}
\begin{tabular}{|c|c|c|c|c|c|c|c||c|c|}
\hline
{}&{}&$A_{i}$&$A_{g}$&$A_{\pi q\bar q}$&$A_{\gamma\pi\pi}$&$A_{c}$&
$A_{\sigma}$&$A_{tot}$&exp. \cite{PDG96}\\
\hline
D$_{13}$(1520)&$A_{1/2}^p(E1)$&+19&+6&--9&--1&+11&--2&+24&{}\\
{}&$A_{1/2}^p(M2)$&--63&--11&0&0&+34&--4&--44&{}\\
{}&$A_{1/2}^p$(tot)&--44&--5&--9&--1&+45&--6&--20&--24$\pm$9\\
\cline{2-10}
{}&$A_{1/2}^n(E1)$&--33&--8&+9&+1&--4&+1&--34&{}\\
{}&$A_{1/2}^n(M2)$&+21&+5&0&0&--11&+1&+16&{}\\
{}&$A_{1/2}^n$(tot)&--12&--3&+9&+1&--15&+2&--18&--59$\pm$9\\
\cline{2-10}
{}&$A_{3/2}^p(E1)$&+32&+10&--16&--2&+19&--2&+41&{}\\
{}&$A_{3/2}^p(M2)$&+37&+6&0&0&--19&+2&+26&{}\\
{}&$A_{3/2}^p$(tot)&+69&+16&--16&--2&0&$0$&+67&+166$\pm$5\\
\cline{2-10}
{}&$A_{3/2}^n(E1)$&--57&--13&+16&+2&--6&+1&--57&{}\\
{}&$A_{3/2}^n(M2)$&--12&--3&0&0&+6&--1&--10&{}\\
{}&$A_{3/2}^n$(tot)&--69&--16&+16&+2&0&$0$&--67&--139$\pm$11\\
\hline\hline
S$_{11}$(1535)&$A_{1/2}^p$(E1)&+114&
--16&--13&-2&--31&+4&+56&+70$\pm$12\\
{}&$A_{1/2}^n$(E1)&--74&+20&+13&+2&+10&--1&--30&--46$\pm$27\\
\hline
\end{tabular}\\[2pt]
\begin{minipage}[b]{13.5cm}
{\small Table 3: Helicity amplitudes for the $\gamma N\rightarrow D_{13}(1520)$ and
$S_{11}(1535)$ transitions in units
of GeV$^{-1/2}\times 10^{-3}$ with the parameters of Table 1
evaluated at $|{\bf q}_{c.m.}|=\omega_{c.m.}$. $A_{i}$ = impulse, 
$A_{g}$ = gluon, 
$A_{\pi q\bar q}$ = pion pair, $A_{\gamma\pi\pi}$ = pion-in-flight, $A_{c}$ = 
confinement, and $A_{\sigma}$ = sigma meson contributions. $A_{tot}$ is 
the sum of all contributions. The E1 and M2 contributions are 
given separately.}
\end{minipage}\\[24pt]

\end{center}

\begin{figure}[htb]
\label{Fig.1}
$$\hspace{0.2cm} \mbox{
\epsfxsize 18.0 true cm
\epsfysize 4.75 true cm
\setbox0= \vbox{
\hbox {
\epsfbox{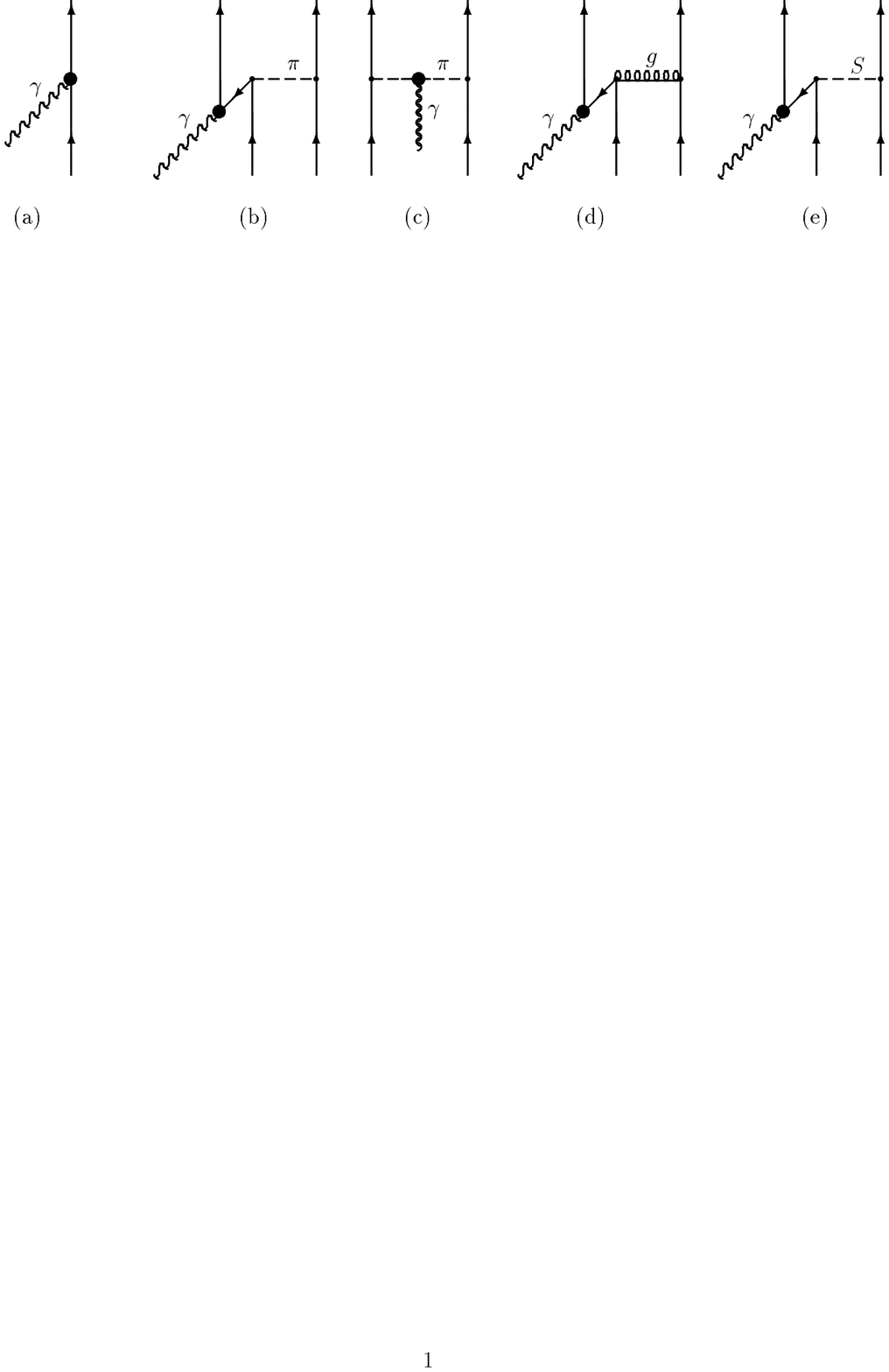}
} 
} 
\box0} $$
\end{figure}
Figure 1: One-body and two-body exchange currents 
between quarks: (a) impulse, (b) pion pair, (c) pion-in-flight, (d)
gluon pair, (e) scalar pair, i.e.\ $\sigma$ meson or confinement pair. 
\noindent\\[24pt]

\end{document}